\documentclass[10pt, conference, compsocconf]{IEEEtran}
\usepackage{afterpage}
\usepackage{xcolor}
\usepackage{graphicx}
\usepackage{float}
\usepackage{longtable}
\usepackage{hyperref}
\hypersetup{
    colorlinks=true,
    linkcolor=blue,
    filecolor=magenta,      
    urlcolor=cyan,
}

\ifCLASSINFOpdf
\else
\fi
\begin{document}
\title{3D Bounding Box Detection in Volumetric Medical Image Data: A Systematic Literature Review }


\author{\IEEEauthorblockN{Daria Kern (Author)}
\IEEEauthorblockA{Faculty of Optics \& Mechatronic\\
Aalen University\\
Aalen, Germany\\
daria.kern@hs-aalen.de, hello@dariakern.de}
\and
\IEEEauthorblockN{André Mastmeyer (Author)}
\IEEEauthorblockA{Faculty of Optics \& Mechatronic\\
Aalen University\\
Aalen, Germany\\
andre.mastmeyer@hs-aalen.de}
}

%Telephone: (800) 555--1212, Fax: (888) 555--1212}

\maketitle

\begin{abstract}
This paper discusses current methods and trends for 3D bounding box detection in volumetric medical image data. For this purpose, an overview of relevant papers from recent years is given. 2D and 3D implementations are discussed and compared. Multiple identified approaches for localizing anatomical structures are presented. The results show that most research recently focuses on Deep Learning methods, such as Convolutional Neural Networks vs. methods with manual feature engineering, e.g. Random-Regression-Forests. An overview of bounding box detection options is presented and helps researchers to select the most promising approach for their target objects.

\end{abstract}


\IEEEpeerreviewmaketitle

\section{Introduction}
The extraction of a Volume of Interest (VOI) is an important pre-processing step in computer based diagnosis. Tasks such as organ segmentation or classification of malignant tumors usually require a prior localization of the corresponding organ or structure. Especially the semantic segmentation of small organs benefits from a preceding localization. By limiting the data to be examined to a VOI, it is ensured that only relevant areas need to be processed and the computing and memory effort is reduced. 
For instance in the field of intervention training and planning, 4D Virtual Reality (VR) simulations require realistic 3D patient organ models in order to be an adequate preparation for training and planning medical procedures \cite{fortmeier2015direct,fortmeier2014virtual}. The automatic reconstruction of such 3D organ models benefits from the localization of a VOI, as it excludes irrelevant regions and therefore making the segmentation of the relevant structures easier and more efficient.
% benefits of VOI:  focus on  regions  that  are  more  likely  to  contain  the  target  organs, improves the computation and memory efficiency, reduces  the  risk  of  false  positive  segmentations

In this Paper we review 3D Bounding Box (BB) detection in volumetric medical image data. Such data is generated by imaging procedures such as CT (Computerized Tomography), MRI (Magnetic Resonance Imaging), PET (Positron Emission Tomography), US (Ultrasound), HFU (High Frequency Ultrasound), just to name a few. We focus only on recently published papers (last five years) to capture current trends and developments.

\section{Methodology}
%Literature search strategy and Quality criteria for study identification
The papers of interest deal with methods to detect 3D BBs around targets in volumetric medical image data. Therefore we used search terms containing \textit{"3D Bounding Box" AND "localization" AND medical -vehicle -"point cloud"} (excluding terms \textit{"vehicle"} and \textit{"point cloud"}) to find relevant papers in public databases and digital libraries. The platforms searched were, IEEE Xplore\footnote{\href{https://ieeexplore.ieee.org/search/searchresult.jsp?queryText=(((((\%22All\%20Metadata\%22:3D\%20Bounding\%20Box)\%20AND\%20\%22All\%20Metadata\%22:localization)\%20AND\%20\%22All\%20Metadata\%22:medical)\%20NOT\%20\%22All\%20Metadata\%22:vehicle)\%20NOT\%20\%22All\%20Metadata\%22:point\%20cloud)&highlight=true&returnFacets=ALL&returnType=SEARCH&matchPubs=true&ranges=2015_2019_Year}{ieeexplore.ieee.org}}, ACM\footnote{\href{https://dl.acm.org/action/doSearch?fillQuickSearch=false&expand=dl&AfterYear=2015&BeforeYear=2020&AllField=\%223D+Bounding+Box\%22+AND+\%22localization\%22+AND+medical+NOT+vehicle+NOT+\%22point+cloud\%22}{dl.acm.org}}, Springer\footnote{\href{https://link.springer.com/search?date-facet-mode=between&facet-content-type=\%22ConferencePaper\%22&showAll=true&query=\%223D+Bounding+Box\%22+AND+\%22localization\%22+AND+medical+NOT+vehicle+NOT+\%22point+cloud\%22+&facet-start-year=2015&facet-end-year=2020}{link.springer.com}}, Google Scholar\footnote{\href{https://scholar.google.de/scholar?q=\%223D+Bounding+Box\%22+AND+\%22localization\%22+AND+medical+-vehicle+-\%22point+cloud\%22&hl=en&as_sdt=0\%2C5&as_ylo=2015&as_yhi=2020}{scholar.google.de}} and WoS\footnote{\href{https://webofknowledge.com}{webofknowledge.com}, "3D Bounding Box"  AND "localization"  AND medical  NOT vehicle  NOT "point cloud",
Timespan: Last 5 years.}. The search was always limited to publications from 2015 to 2020. All papers selected for this review are written in English and have been published internationally. By abstract screening, a total of 31 papers was selected.

\section{3D Bounding Box Representations}
\begin{figure}[hbt]
 \centering
 \frame{\includegraphics[trim={-1cm -1cm -1cm -1cm},clip, scale=0.32]{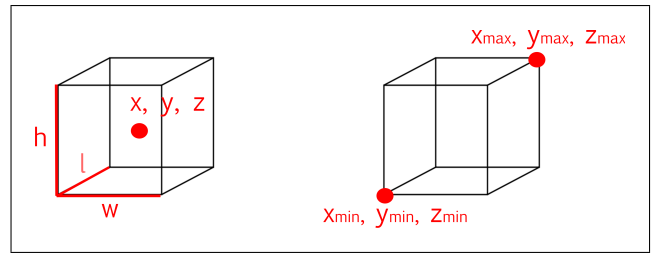}}
 \caption{Possible BB representations. Using centroids (left) or two opposite corners (right).}
 \label{fig:BBrep}
\end{figure}

A 3D BB describes a cuboid object in 3D space. 3D BBs can be represented in different ways. Two common kinds are the centroid and the two corner representations as seen in Fig. \ref{fig:BBrep}. The former defines the center coordinates and the height, width and length of the BB. In the latter case the BB is defined by two opposite corners. Two opposite corners are e.g. the minimum and the maximum coordinate points.

\section{2D vs. 3D Implementation}
In the past, a popular approach was to train a model using handcrafted features. In 2010 Criminisi \cite{RFCriminisi2010} proposed Random Regression Forests (RRF) to localize target structures in 3D Volumes. Unlike traditional approaches, modern Deep Leaning methods like Convolutional Neural Networks (CNN) do not have to rely on handcrafted features, but benefit from automated feature extraction. In recent years the focus has clearly shifted towards Deep Learning. 

\begin{figure}[H]
 \centering
 \frame{\includegraphics[trim={0cm 0cm 0cm 0cm},clip, scale=0.55]{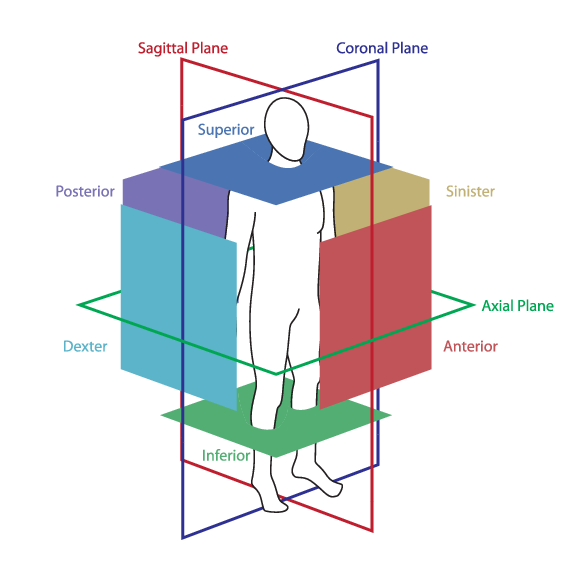}}
 \caption{ BB walls (6 opaque squares). Sagittal, coronal and axial image viewing planes (outlined rectangles) \cite{22BobDeVos2017}.}
 \label{fig:22BobDeVos2017Fig2}
\end{figure}

The implementation of solutions for finding 3D BB for target structures in volumetric data can be performed in 2D or 3D. While a 3D implementation takes the whole volume into account, a 2D implementation distinguishes between three orthogonal image planes. These planes are shown in Fig. \ref{fig:22BobDeVos2017Fig2} as red (sagittal), blue (coronal), green (axial) outlined rectangles. In Fig. \ref{fig:22BobDeVos2017Fig2}, a 3D BB then is constructed by shifting the colored planes (plane/outside normals pointing away from the patient) around a structure of the human body, e.g. the head.

\subsection{Fully 3D Implementations}
The 3D implementation approach takes the whole 3D image volume as an input to detect a 3D BB. 3D CNNs use 3D instead of 2D filter kernels. The 3D kernel has to convolve over three axes, thus capturing context information between slices, but also requiring far more resources than its 2D counterpart. Recent work has made extensive use of 3D CNNs \cite{37Valindria2018, 28Janssens2018, 31Iyer2018, 34Xu2019, 4Wei2019, 1fern2020, 12Han2020, 6Iyer2020}. 3D versions of Deep Learning architectures like VGGNet(\cite{VGGNetSimonyan2014}) \cite{23Qiu2018}, Faster R-CNN(\cite{FasterR-CNNRen2017}) \cite{26Xu2019,10Kaluva2020} and V-Net (\cite{V-NetMilletari2016}) \cite{25Xu2020, 29Zheng2020} are very popular. Although most approaches today rely on CNNs, more traditional approaches are still present. Y. Zhang et al. (2017) \cite{5Zhang2017} train a Random Forest after extracting Haar-like features for every voxel to determine a rough 3D BB and R. Gauriau et al. \cite{33Gauriau2015} use a cascade of two RRFs.

Although comparisons have shown that 3D approaches generally deliver better results \cite{3Dbetter1, 3Dbetter2, 3Dbetter3}, they still come at a cost. The processing in 3D manner requires far more computational resources. The advantage of capturing spatial information in all dimensions goes hand in hand with higher memory demand and required computing power. Furthermore, 3D training data is often not available to the same extent as 2D training data. 

\subsection{2D and 2.5D Implementations} 
The 2D implementation approach deals with 3D localization as a 2D problem. Therefore the volumetric data is examined slice wise in one of the three orthogonal image planes (i.e. sagittal, coronal and axial). The 3D image is thus treated as a stack of several 2D images. A common approach is to use a single 2D CNN or a combination (2.5D) of several (usually three) 2D CNNs for slice wise detection in either one or all three orthogonal viewing plane directions.

\begin{figure}[hbt]
 \centering
 \frame{\includegraphics[trim={0cm 0cm 0cm 0cm},clip, scale=0.485]{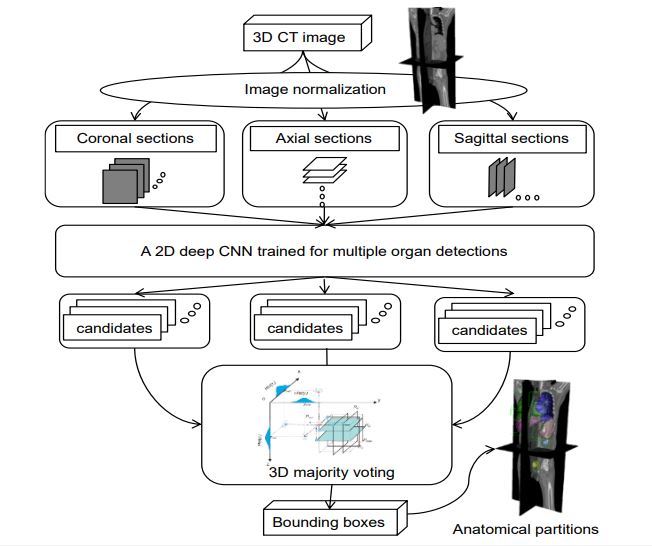}}
 \caption{Exemplary process flow for a single 2D CNN combining 3 orthogonal image plane stacks \cite{11Xiangrong2019}.}
 \label{fig:11Xiangrong2019Fig1}
\end{figure}

A single 2D CNN can be implemented to analyze exactly one of the three image plane stacks \cite{ 17Afshari2018, 21Yang2019, 30Wang2019, 36Tang2018}. Adjacent slices as additional channels \cite{16Lou2019} or dimensions \cite{7Jiang2019} help to capture contextual information. Another possibility is to analyze all three image planes by using a single 2D CNN three times \cite{8Ebner2020,9Ebner2018, 11Xiangrong2019, 22BobDeVos2017} or three separate 2D CNNs per plane \cite{3BobVos2016, 13Wolterink2016, 19Zreik2016, 38Roth2018, 39Huang2018}. Adjacent slices and separate CNNs can also be used in combination \cite{27HumpireMamani2018}. After one or more 2D models have processed the data for multiple slicing directions, the results still have to be combined to create a 3D BB. This can be done by means of a majority voting as seen in Fig. \ref{fig:11Xiangrong2019Fig1}.
In the illustrated workflow, the 3D input image is sliced in all three viewing plane directions. A single 2D model processes the input for each direction separately. The output are three different BBs for the target structure. The coordinates of the BBs are evaluated together and a majority vote determines the final BB.

\begin{figure}[hbt!]
 \centering
 \frame{\includegraphics[trim={0cm 0cm 0cm 0cm},clip, scale=1.45]{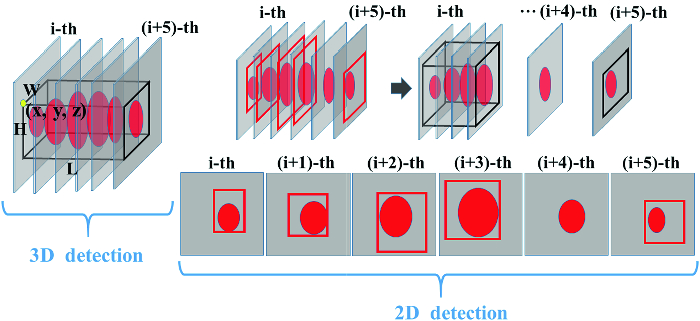}}
 \caption{The problem of 2D detection when assembling the slices to form a cuboid BB \cite{4Wei2019}.}
 \label{fig:4Wei2019Fig1}
\end{figure}

The advantage of a 2D compared to a 3D approach, is the lower memory consumption and the larger amount of training data that results from splitting the 3D images into stacks of several slices. A disadvantage is that context information is usually lost. Furthermore, the results of all slices must be assembled to form a cuboid BB, which is further complicated by occurring spatial discontinuity of the slices as seen in Fig. \ref{fig:4Wei2019Fig1}. In a 3D detection the image is viewed as a whole. The resulting BB therefore seamlessly encloses the target structure. The problem with 2D detection is that the 3D image is broken down into individual sectional images and BBs are determined individually for each image. %It is possible that no BB is detected for individual 2D slices. Resulting in discontinuities in the assembled 3D BB.

\section{Approaches}
The following approaches for 3D BB detection in volumetric medical image data have been identified amongst the investigated papers.

\subsection{Slice Wise Box Detection}
%slice wise presence prediction for every plane
This approach simply detects the presence of the target structure in every slice. The results for each orthogonal image plane stack are combined to produce a 3D BB \cite{3BobVos2016, 13Wolterink2016, 19Zreik2016, 22BobDeVos2017}. The approach works regardless of whether the results were generated by a single 2D CNN or a combination of three 2D CNN.

\subsection{Coarse Segmentation / Probability Maps}
The coarse-segmentation of target structures is often an intermediate step for a subsequent refined segmentation. First, the entire image is viewed to roughly locate one or more targets. The resulting sub-optimal segmentation is then utilized to place a BB around the area of interest \cite{7Jiang2019, 8Ebner2020, 9Ebner2018, 30Wang2019, 37Valindria2018,  16Lou2019, 25Xu2020, 29Zheng2020, 39Huang2018}.

\begin{figure}[H]
 \centering
 \frame{\includegraphics[trim={0.75cm 0.75cm 0cm -0.25cm},clip, scale=0.255]{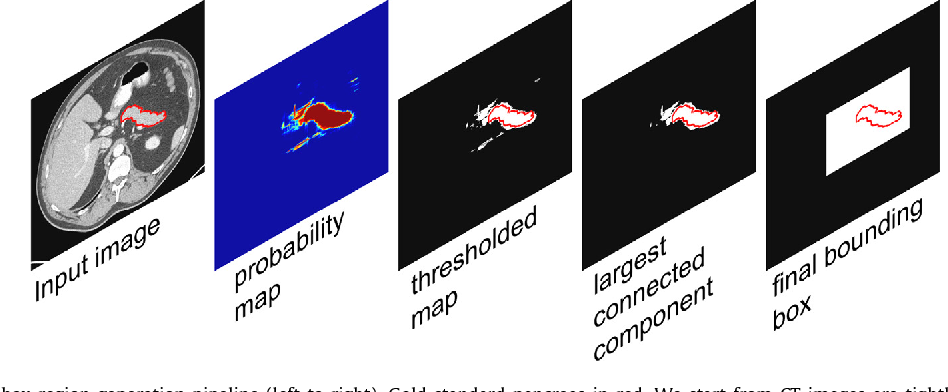}}
 \caption{Procedure implemented by H. Roth et al. (2018) \cite{38Roth2018}.}
 \label{fig:38Roth2018Fig3}
\end{figure}

Similar to a coarse-segmentation approach,
H. Roth et al. (2018) \cite{38Roth2018} implement a 2D pixel-wise probability detection in every image plane direction to obtain confidence heat-maps, which are then used to generate a 3D BB. By applying a threshold against the pixel probabilities, the largest connected component is found and a BB is simply put around it. The procedure is shown step by step in Fig.\ref{fig:38Roth2018Fig3}. R. Gauriau et al. (2015) \cite{33Gauriau2015} calculate voxel probabilities to obtain confidence maps in a 3D manner. They utilize RRFs and divide the localization into 2 steps. A first RRF performs a rough localization of all organs at once. A second, organ-specific RRF focuses on the individual organs respectively. In a similar fashion Y. Zhang et al. (2017) \cite{5Zhang2017} first take advantage of the knowledge about the relative positions of the target structures and their voxel intensity by using haar-like features to narrow down the target area. A RRF is then trained on spatial and intensity features to predict a voxel-wise probability map within the target area. Using a threshold, a BB is placed around the target structure.

\subsection{Deep Reinforcement Learning}
Deep Reinforcement Learning (DRL) combines Reinforcement Learning (RL) and Deep Learning. In RL an agent takes a sequence of actions in order to achieve a certain goal. In doing so, it receives feedback in form of rewards and penalties. Through trial and error, the agent tries to maximize the accumulated reward and learns which actions to take. DRL incorporates Deep Neural Networks (DNN) into this task. The DNN analyzes the current state and decides which action to take. In the work of F. Navarro et al. (2020) \cite{1fern2020}, the CNN receives the current BB voxel values and those of the last four states as input for performing the task of finding the final BB. The actions consider the moving direction, translation and scaling of the 3D BB. S. Iyer et al. (2018, 2020) \cite{31Iyer2018, 6Iyer2020} employ two 3D CNNs, one for learning the navigation in the coordinate directions and the other to predict the size of the BB dimensions. 
%deep Q \cite{DeepQMnih2015} 

\subsection{Anchor Based Approaches}
Another often seen approach is using anchor boxes, which are predefined BB guesses of certain scales and aspect ratios. For instance M. Tang et al. (2018) \cite{36Tang2018} and Y. Wei et al. (2019) \cite{4Wei2019} follow this approach. Latter combine a 3D CNN and an additional 2D feature extractor for the axial slice direction to handle various scales and shapes of the target structure. An output predictor takes the resulting features as input. Very popular anchor-based approaches are Faster R-CNN \cite{FasterR-CNNRen2017} and YOLO \cite{YOLORedmon2017}. S. Afshari et al. (2018) \cite{17Afshari2018} use a modified 2D YOLO to analyze the coronal image plane stack. Whereas YOLO is a one-stage detector, the Faster R-CNN workflow consists of two stages. The backbone network extracts features, which are, together with the anchor boxes, used by a Region Proposal Network (RPN) to generate BB candidates. A Fast R-CNN \cite{Fast-RCNNGirshick2015} classifier and regressor are then used to determine the class of the object and refine the BBs. K. Chaitanya et al. (2020) \cite{10Kaluva2020} and  X. Yang et al. (2019) \cite{21Yang2019} use a 3D and 2D Faster R-CNN architecture respectively to detect BBes. X. Xu et al. (2019b) \cite{26Xu2019} modify the 3D Faster R-CNN architecture by removing the classifier and using the Region Proposal Network to propose organ-specific BBs. Relying on the fact that there is at most one instance of a organ, BBs with the same label are merged into one.

\begin{figure}[hbt]
 \centering
 \frame{\includegraphics[trim={0cm 0cm 0cm 0cm},clip, scale=0.155]{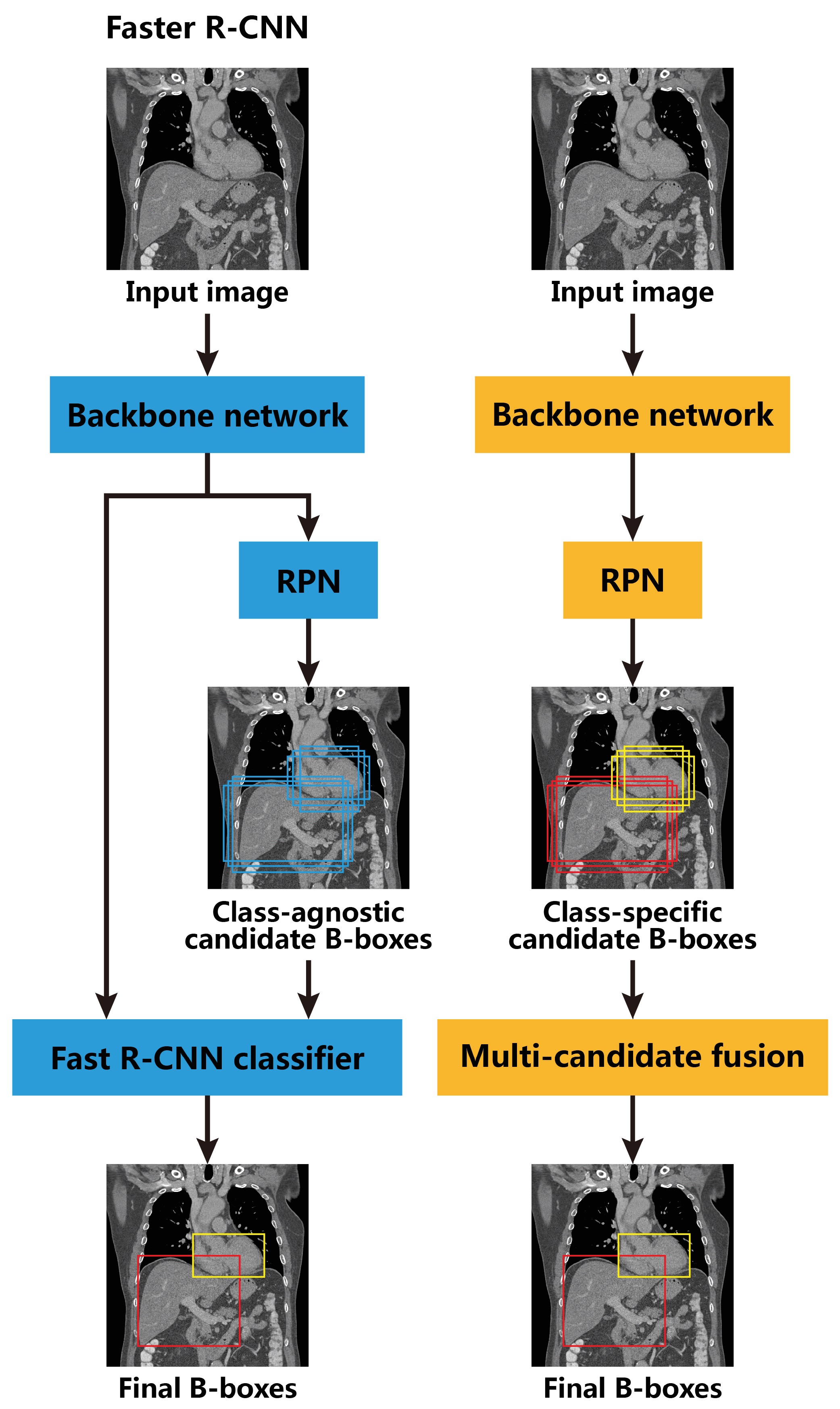}}
 \caption{left: general Faster-R-CNN \cite{FasterR-CNNRen2017} workflow, right: X. Xu et al. (2019b) \cite{26Xu2019} workflow.}
 \label{fig:26Xu2019Fig1}
\end{figure}

Fig. \ref{fig:26Xu2019Fig1} illustrates the common workflow and the one adapted by X. Xu et al. (2019b) \cite{26Xu2019}. L. Liu et al. (2019) \cite{35Lui2019} first identify target regions with a Conditional Gaussian Model (CGM) and further localize target structures using a 2D Faster R-CNN.

\subsection{Other Approaches}
S. Han et al. (2020) \cite{12Han2020} use a 3D modified pre-activation ResNet \cite{ResNetHe2016} for regression on the BB coordinates. R. Janssens et al. (2018) \cite{28Janssens2018} also use regression to predict two relative displacement vectors between the two  diagonal corners of a BB and a  reference voxel. 

Z. Qiu et al. (2018) \cite{23Qiu2018} scan the whole volume using a 3D sliding window, that is large enough to fully contain the target structure. A 10-layer VGGNet \cite{VGGNetSimonyan2014} serves as the classifier.

X. Xu et al. (2019a) \cite{34Xu2019} binarize the predicted sagittal, axial and coronal presence probability curves of the target organs by applying a threshold. The 3D BBs are composed by the largest 1D nonzero component in these three binary curves.

\section{Results}
Table \ref{tab:overview} gives an overview of the evaluated papers. Included are the author, the image modality, the approach to 3D BB detection, the target structure in the body and the evaluation results of the work. The "Results" column in Table \ref{tab:overview} is nonexhaustive. B. de Vos et al. (2017) \cite{22BobDeVos2017}, for instance, did extensive testing and a more detailed evaluation can be found in their paper. Some results are also left blank, since no evaluation was performed as localization was a less important intermediate step in these papers. Measured was mostly Intersection over Union (IoU), Dice Similarity Coefficient (Dice), Average Precision (AP) and Wall Distance (WD).

%evtl Which methods achieve the best results R. Huang et al. (2018) \cite{39Huang2018} real-time solution
%B. de Vos et al. (2017) \cite{22BobDeVos2017} did extensive testig/evaluation (mehr dazu ausführlich schrieben vielleicht)

\clearpage
\onecolumn
\begin{longtable}{p{2.3cm}|p{0.55cm}|p{6cm}|p{1.7cm}|p{4.75cm}}
\caption{Literature for 3D BB detection. IoU: Intersection over Union. }
\label{tab:overview}\\
\hline
%\centering
\textbf{Author} & \textbf{Data} & \textbf{Approach} & \textbf{Target(s)} & \textbf{Results}\\

\hline R. Gauriau et al. (2015) \cite{33Gauriau2015}
& CT
& two cascaded RRF. 1st RRF for global coarse segmentation and 2nd organ specific RRF for local BB improvement
& 6 abdominal organs
& mean WD $10.7\pm4$ mm, $5.5\pm4$ mm, $5.6\pm3$ mm, $7.9\pm4$ mm, $9.5\pm4$ mm, $13.2\pm5$ mm\\

\hline B. de Vos et al. (2016) \cite{3BobVos2016}
& CT
& combination of three 2D CNNs (AlexNet \cite{AlexNetKrizhevsky2012}), each analyzing one orthogonal image plane stack
& heart, aortic arch, d. aorta
& median Dice: 0.89, 0.70, 0.85\\

\hline M. Zreik et al. (2016) \cite{19Zreik2016}
& CT
& %combination of three 2D CNNs (AlexNet \cite{AlexNetKrizhevsky2012}), each analyzing one orthogonal image plane stack,
see Bob D. de Vos et al. (2016) \cite{3BobVos2016}
& left ventricle
& complete left ventricle was contained within the BB in all test scans\\

\hline J. Wolterink et al. (2016) \cite{13Wolterink2016}
& CT
& %combination of three 2D CNNs (AlexNet \cite{AlexNetKrizhevsky2012}), each analyzing one orthogonal image plane stack,
see Bob D. de Vos et al. (2016) \cite{3BobVos2016}
& heart
& in all cases the BB contained the whole heart\\

\hline B. de Vos et al. (2017) \cite{22BobDeVos2017}
& CT
& single  2D  CNN (comparing BoBNet \cite{22BobDeVos2017}, VGGNet-16 \cite{VGGNetSimonyan2014}, ResNet-34 \cite{ResNetHe2016} and AlexNet \cite{AlexNetKrizhevsky2012}) analyzes  all three orthogonal image plane stacks
& liver,  heart, a. aorta, aortic arch, d. aorta
& Dice (comparing CNNs) 0.967, 0.963, 0.960, 0.959, WD (BoBNet for liver and heart) $8.87\pm15.00$ mm, $3.11\pm3.43$ mm\\

\hline Y. Zhang et al. (2017) \cite{5Zhang2017}
& CT
& Combination of 3D Haar-like feature \cite{HaarViola2001} extraction for every voxel and a RF
& l.\&r. lung, heart
& /\\

\hline H. Roth et al. (2018) \cite{38Roth2018}
& CT
& combination of three 2D CNNs (HNN \cite{HNNXie2015}), each analyzing one orthogonal image plane stack
& Pancreas
& BB completely surround the pancreases with nearly 100\% recall\\

\hline V. Valindria et al. (2018) \cite{37Valindria2018}
& MRI
& weighted 3D CNN for coarse segmentation, using larger weights for smaller organs  %\textcolor{red}{multi-atlas technique was chosen for producing spatial priors?}
& 11 abdominal organs, 7 bones
& /\\

\hline M. Tang et al. (2018) \cite{36Tang2018}
& US
& single 2D CNN (VGGNet-16 \cite{VGGNetSimonyan2014}) analyzes one orthogonal image plane stack
& femoral head
& /\\

\hline R. Huang et al. (2018) \cite{39Huang2018}
& US
& combination of three 2D CNNs (View-based Projection Networks (VP-Nets)), each analyzing one orthogonal image plane stack in real-time
& 5 key brain structures
& center deviation: $1.8\pm1.4$ mm, size difference: $1.9\pm1.5$ mm, 3D IoU: $63.2\pm14.7$\%\\

%\hline C. O. Laura et al. (2019) \cite{15Laura2019}
%& CT
%& single 2D CNN (modified YOLO v2 \cite{YOLORedmon2017}) analyzes one orthogonal image plane stack, predicting an irregular polyhedron instead of a BB
%& nasal cavity and paranasal sinuses
%& ...\\

\hline S. Afshari et al. (2018) \cite{17Afshari2018}
& PET
& single 2D CNN (modified YOLO \cite{YOLORedmon2017}) analyzes coronal image plane stack
& brain, heart, bladder, r.\&l. kidney
& avg. precision 75-98\%, recall  94-100\%, centroid dist. $<14$ mm, WD $<24$ mm\\

\hline Z. Qiu et al. (2018) \cite{23Qiu2018}
& HFU
& 3D CNN (10-layer VGGNet \cite{VGGNetSimonyan2014})
& brain verticle %(mouse embryos)
& BB containing entire brain verticle 93.7\% (single classifier), 96.4\% (ensemble of 3 classifiers)\\

\hline G. Humpire-Mamani et al. (2018) \cite{27HumpireMamani2018}
& CT
& combination of three 2.5D (adjacent slices) CNNs, each analyzing  one orthogonal image plane stack
& 11 thorax-abdomen organs
& avg. WD of $3.20\pm7.33$ mm, 2nd human observer achieved $1.23\pm3.39$ mm \\

\hline R. Janssens et al. (2018) \cite{28Janssens2018}
& CT
& 3D CNN %(FCN \cite{FCNLong2015})
& lumbar vertebrae
& /\\

\hline S. Iyer et al. (2018) \cite{31Iyer2018}
& CT
& combination of two 3D CNN for Deep Reinforcement Learning and Imitation Learning 
& thoracic \&lumbar vertebrae
& IoU 67.52\%, Dice 80.23\% \\

\hline M. Ebner et al.(2018) \cite{9Ebner2018} and (2020) \cite{8Ebner2020}
& MRI
& single 2D CNN (P-Net \cite{P-NetWang2019}) for coarse segmentation analyzes all three orthogonal image plane stacks
& fetal brain
& IoU 86.54\% (normal), 84.74\% (presurgical), 83.67\% (postsurgical) \\

\hline X. Wang et al. (2019) \cite{30Wang2019}
& US
& single 2D CNN (U-Net \cite{U-NetRonneberger2015}) analyzes one orthogonal image plane stack for coarse segmentation
& fetal femur
& IoU 78.1\%\\

\hline X. Xu et al. (2019a) \cite{34Xu2019}
& CT
& single triple-branch 3D CNN %(FCN \cite{FCNLong2015}) 
with a branch for every orthogonal image plane stack. Additionally creating a three-channel image as input
& 11 body organs
& IoU 76.44, mean WD $4.36\pm7.98$ mm, mean centroid distance $6.91\pm9.66$ mm\\

\hline L. Liu et al. (2019) \cite{35Lui2019}
& PET /CT
& combination a conditional Gaussian model (CGM) and a 2D CNN (Faster R-CNN \cite{FasterR-CNNRen2017}) for refinement, analyzing one orthogonal image plane stack
& heart, liver, spleen, l.\&r. kidney
& centre position error thorax: $7.00\pm2.87$ mm (CT), $4.47\pm2.50$ mm (PET) Abdomen: $4.72\pm2.23$ mm (CT), $4.41\pm2.02$ mm (PET)\\

\hline X. Xu et al. (2019b) \cite{26Xu2019}
& CT
& 3D CNN (modified Faster R-CNN \cite{FasterR-CNNRen2017})
& 11 body organs, 12 head organs
& body: precision 97.91\%, recall 98.71\%, AP 98.24\%, head: 91.11\% 91.11\%, 84.78\%\\

\hline Y. Wei et al. (2019) \cite{4Wei2019}
& CT
& hybrid multi-atrous and multi-scale network (HMMNet) with multi-atrous 3D CNN (MA3DNet) backbone
& liver lesions
& Dice 54.8\% and 34.2\% with IoU of 0.5 and 0.75 respectively\\

\hline H. Jiang et al. (2019) \cite{7Jiang2019}
& CT
& single 2.5D (5 adjacent slices, 3D Conv-Kernel) Attention Hybrid Connection Network (AHCNet) for coarse segmentation analyzes one orthogonal image plane stack.
& liver
& /\\

\hline X. Zhou et al. (2019)  \cite{11Xiangrong2019}
& CT
& single 2D CNN analyzes all three orthogonal image plane stacks
& 17 torso organs
& Sucessfully localized 84.3\% (IoU $\geq0.5$), mean IoU 70.2\%\\

\hline X. Yang et al. (2019) \cite{21Yang2019}
& MRI
& single 2D CNN (Faster-RCNN \cite{FasterR-CNNRen2017}) analyzes one orthogonal image plane stack
& left atrium region
& 100\% accuracy\\

\hline J. Lou et al. (2019) \cite{16Lou2019}
& MRI
& single 2D (adjacent slices as additional channels) CNN (DS U-net \cite{DSU-NetDou2016}) for  coarse  segmentation analyzes one orthogonal image plane stack
& fetal brain
& IoU $91.31\pm0.08\%$, centroid dist. $2.90\pm3.53$ mm\\

\hline F. Navarro et al. (2020) \cite{1fern2020} 
& CT
& 3D CNN (similar to d DQN-based network architecture \cite{DQNAlansary2019}) for Deep Reinfocement Learning
& 7 abdominal organs
& IoU 0.63, abs. median WD $2.25$ mm, median dist. between centroids $3.65$ mm\\

\hline K. Chaitanya et al. (2020)  \cite{10Kaluva2020}
& CT
& 3D CNN (Faster R-CNN \cite{FasterR-CNNRen2017})
& lung nodules
& sensitivity 93\% (nodules$>5$ mm), 91\% (nodules$>3$ mm) \\

\hline S. Han et al. (2020) \cite{12Han2020}
& MRI
& 3D CNN (modified pre-activation ResNet \cite{PreactivationHe2016})
& cerebellum %(brain)
& /\\

%\hline R. Jurdi et al. (2020) \cite{20ElJurdi2020}
%& CT /MRI
%& 
%& heart, tra-chea,
%& aorta, esophagus
%& \\

\hline T. Xu et al. (2020) \cite{25Xu2020}
& HFU
& 3D CNN (similar to V-Net \cite{V-NetMilletari2016}) for coarse segmentation
& embryonic mice brain ventricle \& body %(mouse embryos)
& Dice (coarse segmentation) 0.818, 0.918 \\

\hline S. Iyer et al. (2020) \cite{6Iyer2020}
& CT
& %combination of two 3D CNN for Deep Reinforcement Learning and Imitation Learning, 
see S.  Iyer  et al. (2018) \cite{31Iyer2018}
& thoracic \&lumbar vertebrae
& IoU 74/85\% (chest), Dice 77/86\% (abdomen)\\

\hline H. Zheng et al. (2020) \cite{29Zheng2020}
& CT
& two cascaded 3D  CNN  (V-Net\cite{V-NetMilletari2016}) for coarse segmentation
& pancreas
& 1st\&2nd V-Net Dice: $81.38\pm6.48\%$, $81.79\pm7.10\%$, sensitivity  $80.55\pm9.36\%$, $81.51\pm7.22\%$ \\
\hline
\end{longtable}
%\clearpage
\twocolumn
%\afterpage{\blankpage}

\section{Conclusion and Future Work}

We provide a synopsis of the recent works dealing with 3D BB detection in volumetric medical images. For this purpose 31 papers of the last 5 years were evaluated. The review is intended to provide an overview of the current trends as well as information on various options for BB detection in 3D data. 3D and 2D implementations were differentiated, processing the 3D input as a whole or splitting it into several 2D inputs. Various approaches were identified, Coarse Segmentation being the most commonly used. It was also found that Deep Learning methods have largely replaced traditional and other methods, e.g. RRF. The overview of options presented in this review will help future researchers to select a promising approach, which also reflects the state of research.
%Ultimately, there are many ways to reach the goal. Future Research could investigate the applicability of the identified options in various cases. 
Some of the presented techniques are also applicable to 2D imagery, e.g. detecting, learning and discerning face appearances in photographs \cite{Kern2020mayibuye}. Traditional techniques such as RRFs have been augmented by Deep Learning techniques, especially with CNNs among them. The most promising and increasingly successful methods seem to be CNNs, as they combine traditional signal processing approaches (convolution filtering) with automatic learning from examples in Neural Networks. BB detection helps to save computational cost and to train models for the subsequent semantic segmentation of body areas more specifically, with better results in the end.

To assess the quality and relevance of BB detection for patient modelling in VR simulators \cite{mastmeyer2016efficient, fortmeier2015direct, fortmeier2014virtual} thoroughly, we plan studies in our lab to examine the influence of different imaging modalities \cite{zaffino2019fully,mastmeyer2017accurate,mastmeyer2016real,mastmeyer2016random,mastmeyer2015model} and BB detection quality by VR visualization and interaction with detected BBs using haptic force feedback \cite{mastmeyer2017evaluation,mastmeyer2016random,mastmeyer2014ray,fortmeier2012gpu,mastmeyer2012direct} for quality assurance. In the future, we will also address the accurate and precise BB detection and content segmentation \cite{mastmeyer2013patch} using nD image data from various imaging sources. Additionally the quality of organ models in the time-dynamic simulation of 4D medical needle \cite{mastmeyer2013ray,fortmeier2013optimized} interventions \cite{mastmeyer2018population,mastmeyer2017interpatient} shall profit from the hierarchical and more specific approach.

\section*{Acknowledgment}
German Research Foundation DFG MA 6791/1-1, EXPLOR-19AM funds granted by Foundation Kessler+Co. for Education and Research. %NVIDIA GPU grant 2018.

% References 
\bibliography{bare_conf} % bibliography data in bib.bib
\bibliographystyle{IEEEtran} % makes bibtex use IEEEtran.bst

\end{document}